\documentclass[natbib]{emulateapj}
\usepackage{natbib}
\usepackage{bm}
\def\s2n{S^{\prime}/N}

\def\gsim{\;\rlap{\lower 2.5pt
\hbox{$\sim$}}\raise 1.5pt\hbox{$>$}\;}
\def\lsim{\;\rlap{\lower 2.5pt
\hbox{$\sim$}}\raise 1.5pt\hbox{$<$}\;}

\usepackage{amssymb,amsmath}
\def\bs{\boldsymbol}

\slugcomment{Submitted to ApJ, \today}
\shorttitle{Exact Results in Supersonic Turbulence}
\shortauthors{Pan et al.}

\begin{document}
\title{Detailed Balance and Exact Results for Density Fluctuations in Supersonic Turbulence}

\author{Liubin Pan,}
\affil{School of Physics and Astronomy, Sun Yat-sen University, 2 Daxue Road, Zhuhai, Guangdong, 519082, China; panlb5@mail.sysu.edy.cn}
\author{Paolo Padoan,}
\affil{Institut de Ci\`{e}ncies del Cosmos, Universitat de Barcelona, IEEC-UB, Mart\'{i} Franqu\`{e}s 1, E08028 Barcelona, Spain; 
ppadoan@icc.ub.edu}
\affil{ICREA, Pg. Llu\'{i}s Companys 23, 08010 Barcelona, Spain}
\author{{\AA}ke Nordlund}
\affil{Centre for Star and Planet Formation, Niels Bohr Institute and Natural History Museum of Denmark, University of Copenhagen, {\O}ster Voldgade 5-7, DK-1350 Copenhagen K, Denmark; aake@nbi.ku.dk}

\begin{abstract}
The probabilistic approach to turbulence is applied to investigate density fluctuations in supersonic turbulence. We derive kinetic equations for the probability distribution 
function (PDF) of the logarithm of the density field, $s$, in compressible turbulence in two forms: a first-order partial differential equation involving 
the average divergence conditioned on the flow density, $\langle \nabla \cdot {\bs u} | s\rangle$, and a Fokker-Planck equation with the drift and diffusion 
coefficients equal to $-\langle  {\bs u} \cdot   \nabla s | s\rangle$ and $\langle  {\bs u} \cdot   \nabla s | s\rangle$, 
respectively. Assuming statistical homogeneity only, the detailed balance at steady state leads to two 
exact results, $\langle \nabla \cdot {\bs u} | s \rangle =0$, and $\langle  {\bs u} \cdot  \nabla s | s\rangle=0$. The former indicates a balance of the flow divergence 
over all expanding and contracting regions at each given density. 
The exact results provide an objective criterion to judge the accuracy of numerical codes with respect to the density statistics in supersonic turbulence. 
We also present a method to estimate the effective numerical diffusion as a function of the flow density  and discuss its effects on the shape of the density PDF. 
\end{abstract}


\section{Introduction}
Supersonic turbulence in molecular clouds plays a crucial role in the 
process of star formation. The probability distribution function (PDF) of density fluctuations in supersonic 
turbulence has been extensively investigated \citep[e.g.][]{Vazquez-Semadeni94,Padoan+97,Nordlund+Padoan99,Molina+12} and 
widely used in theoretical models of star formation \citep{Krumholz+McKee05,Padoan+Nordlund11sfr,Hennebelle+Chabrier11,Federrath+Klessen12}. 
In star formation models based on turbulent fragmentation, the shape of the 
density PDF, particularly its high-density tail, is of particular importance, due to its impact on the 
star formation rate and the predicted stellar initial mass function
\citep[e.g.][]{Padoan+97,Padoan+Nordlund02,Hennebelle+Chabrier08,Padoan+Nordlund11sfr}. 
Numerical simulations of isothermal supersonic turbulence with solenoidal forcing have shown that the density PDF is generally consistent with a 
lognormal distribution, whereas changes in the equation of state \citep{Passot+Vazquez-Semadeni98,Scalo+98}, the forcing pattern 
\citep{Federrath+08,Federrath+10}, and the inclusion of gravity \citep{Collins+11,Kritsuk+11} all induce variations in the PDF shape.   

The density PDFs used in  star-formation models are usually based on results from numerical simulations. 
The theoretical understanding of the origin of such PDFs is still incomplete, with most interpretations 
of numerical results being heuristic or qualitative. For example, the usual argument that the 
log-normal distribution is the consequence of a multiplicative process of successive, independent 
compressions and expansions is purely phenomenological. Also, it is not clear how artificial 
numerical diffusion that exists in all simulations affects the PDF shape. 

In this Letter, we study the density statistics from first principles, by deriving 
kinetic equations of the density PDF. Exact results corresponding to the detailed balance of probability fluxes at 
steady state are derived using the assumption of statistical homogeneity only (\S 2). We stress that, 
due to strong nonlinearity, exact results in turbulence are very rare, with the known examples being Kolmogorov's celebrated 4/5 law 
and similar ones in different flow cases \citep[e.g.][]{Yaglom49, Politano+Pouquet98, Galtier+Banerjee11}.
The exact results are used to test the accuracy of numerical simulations in \S~3, and our
conclusions are summarized in \S 4.

\section{The PDF Equations and Exact Results}

\subsection{The PDF equations}
Defining the logarithm of the density, $s\equiv \ln (\rho/\langle \rho \rangle)$, with $\langle \rho \rangle$ the average 
density, the continuity equation reads,   
\begin{equation}
\frac {\partial s}{\partial t}  + \bs{u}  \cdot \nabla s = - \nabla \cdot \bs{u},
\label{s-eq}
\end{equation}
where $\bs{u}$ is the turbulent velocity. Following the general procedure of the probabilistic approach for 
turbulence studies\citep[e.g.][]{Pope00}, we define a fine-grained PDF, $g(\zeta; {\bs x}, t) = \delta(\zeta - s({\bs x}, t))$, 
where $\delta$ is the Dirac delta function and $\zeta$ the sampling variable. 
The time derivative of $g$ is given by $\partial_t g(\zeta; {\bs x}, t) = -\partial_\zeta g \partial_t s $, as $g$ 
depends on $t$ only through $ (\zeta - s({\bs x}, t))$.  
Using Eq.\ (\ref{s-eq}) for $\partial_t s$ yields,
\begin{equation}
\frac {\partial g (\zeta; {\bs x}, t)}{\partial t}   =    \frac {\partial  (g  \bs{u}  \cdot \nabla s)}{\partial \zeta} + \frac{ \partial (g \nabla \cdot \bs{u}) }{\partial \zeta },
\label{finegrained}
\end{equation}
where the last two terms use the fact that $\nabla \cdot \bs{u}$ and  $\bs{u}  \cdot \nabla s$  are independent of the sampling variable, $\zeta$.

The coarse-grained PDF is defined as the ensemble average of $g$, i.e., $f(\zeta; {\bs x}, t) \equiv \langle g(\zeta; {\bs x}, t) \rangle$, over independent flow realizations.
The ensemble average of the product of any quantity, $\phi ({\bs x}, t)$, with $g$ (a delta function) can be written in terms of 
a conditional average, $\langle  \phi ({\bs x}, t) \delta(\zeta - s({\bs x}, t)) \rangle = \langle  \phi ({\bs x}, t)| s({\bs x}, t) = \zeta \rangle f(\zeta; {\bs x}, t)$, where $\langle  \phi ({\bs x}, t)| s({\bs x}, t) = \zeta \rangle$ is 
the average of $\phi$ over the realizations where $s( {\bs x}, t) $ equals the sampling variable \citep[]{Pope00}. 
Ensemble averaging Eq.\ (\ref{finegrained}) then gives, 
\begin{equation}
\frac {\partial f (\zeta; {\bs x}, t)}{\partial t}   = \frac {\partial} {\partial \zeta } \Big(\langle   \bs{u}  \cdot \nabla s|s=\zeta \rangle f\Big) + \frac {\partial} {\partial \zeta }  \Big(\langle \nabla \cdot \bs{u} |s=\zeta\rangle f\Big),
\label{coarse}
\end{equation} 
where the last two terms represent the fluxes of probability into and out of a given $s$ interval by the advection of  $s$ and the divergence, 
respectively. At steady state, a balance of the probability flux is expected. 
We refer to $\langle \nabla \cdot \bs{u} |s=\zeta\rangle$ and $\langle \bs{u} \cdot \nabla s |s =\zeta\rangle$, as the conditional mean divergence
and conditional mean advection, respectively.

An important relation exists between the two conditional means. 
Ensemble averaging the equality $g\nabla \cdot \bs{u}  = \nabla \cdot ( g \bs{u} ) -  \bs{u} \cdot \nabla g =  \nabla \cdot ( g \bs{u} ) +   (\bs{u} \cdot \nabla s) \partial_\zeta  g$, 
and assuming statistical homogeneity, we find that, 
\begin{equation}
\langle \nabla \cdot \bs{u}| s=\zeta \rangle f = 
\frac{\partial }{\partial \zeta} \Big( \langle \bs{u} \cdot \nabla s |s =\zeta \rangle f \Big).
\label{div-adv-relation}
\end{equation} 
Using this relation in Eq.\ (\ref{coarse}) leads to two forms of kinetic equations for $f(\zeta; {\bs x}, t)$, one of which is, 
\begin{equation}
\frac {\partial f } {\partial t}  = \langle \nabla \cdot \bs{u}| s=\zeta \rangle f   + \frac {\partial}  {\partial \zeta} \Big(\langle \nabla \cdot \bs{u}| s=\zeta \rangle f \Big),
\label{pdf-div}
\end{equation} 
where the PDF evolution is determined by the conditional mean divergence.  The other form is a Fokker-Planck equation,
 \begin{equation}
\frac {\partial f } {\partial t}  =  
\frac {\partial}  {\partial \zeta}  \Big(\langle \bs{u} \cdot \nabla s |s =\zeta \rangle f \Big) + 
\frac {\partial^2}  {\partial \zeta^2}  \Big (\langle \bs{u} \cdot \nabla s |s =\zeta \rangle f \Big),
\label{pdf-adv}
\end{equation} 
where the drift and diffusion coefficients are $-\langle \bs{u} \cdot \nabla s |s =\zeta \rangle$ and $\langle \bs{u} \cdot \nabla s |s =\zeta \rangle$, 
respectively. The diffusion term in Equation (\ref{pdf-adv}) tends to broaden the PDF, while the drift term reduces the mean of $s$. 
By analyzing $\langle \bs{u} \cdot \nabla s |s =\zeta \rangle$, the Fokker-Planck equation may be conveniently 
used to study the development of density fluctuations  and the evolution of the PDF from initial conditions.  

Equations (\ref{pdf-div}) and (\ref{pdf-adv}) for the PDF of $s$ are exact; however, they are not 
closed (hence not directly solvable) because the conditional means involve two-point, 
density-velocity joint statistics, whose evolution relies on three-point joint statistics and so on.    
For simplicity, we will drop the sampling variable, $\zeta$, and write the PDF 
as $f(s; t)$, and any conditional mean $\langle ... |s =\zeta \rangle  $ as $\langle ... |s \rangle$.

\subsection{Exact results at statistically steady state}

At statistically steady state, exact results corresponding to the 
balance of probability fluxes can be derived from the PDF equations.
At steady state, equation (\ref{pdf-div}) is solved by
$\langle \nabla \cdot \bs{u}| s \rangle f (s) = C \exp(-s)$,
where $C$ is the integration constant. The integral of  $\langle \nabla \cdot \bs{u}| s \rangle f (s)$ 
from $-\infty$ to $\infty$ is $\langle \nabla \cdot  \bs{u} \rangle$, which is 0 from homogeneity. 
This requires $C=0$, so that, 
\begin{equation}
\langle \nabla \cdot \bs{u}| s \rangle =0, 
\label{zero-div}
\end{equation}
for all $s$, indicating that, at each given density, the values of the velocity divergence in expanding and 
converging regions of a compressible turbulent flow cancel out exactly.  This exact balance  at 
each density is a {\it detailed}  version of the overall balance, $\langle \nabla \cdot {\bs u} \rangle =0$, that follows simply from statistical homogeneity. 

Combining Equation (\ref{zero-div}) and Equation  (\ref{div-adv-relation}) gives $ \langle \bs{u} \cdot \nabla s |s \rangle f  =C_2$,  with $C_2$ 
another integration constant. Considering that $ \int_{- \infty}^{\infty} \langle \bs{u} \cdot \nabla s |s \rangle f(s) d s =\langle \bs{u} \cdot \nabla s\rangle$ and 
that $\langle \bs{u} \cdot \nabla s\rangle=0$ at steady state (as can be seen by averaging Equation (\ref{s-eq})), we have $C_2=0$ and, 
\begin{equation}
\langle \bs{u} \cdot \nabla s |s  \rangle =0,
\label{zero-adv}
\end{equation}
for all $s$. Note that our main results, Eqs.\ (\ref{zero-div}) and (\ref{zero-adv}), are derived exactly, with only assumptions of statistical 
homogeneity and stationarity. 

Eqs.\ (\ref{zero-div}) and (\ref{zero-adv}) indicate that the probability fluxes due to the advection and divergence terms in 
Eq.\ (\ref{coarse}) are perfectly balanced individually. This individual balance of each term is not required by Eq.\ (\ref{coarse}), which 
only demands an overall balance $\langle \bs{u} \cdot \nabla s |s  \rangle +\langle \nabla \cdot \bs{u}| s \rangle =0$ at 
steady state.  It is the relation in Eq.\ (\ref{div-adv-relation}) that leads to individual balances of the two terms. 
We will refer to both Eq.\ (\ref{zero-div}) and Eq.\ (\ref{zero-adv}) as detailed balance. 
 
\section{Simulation results}

\subsection{Effects of artificial numerical diffusion}

The results derived in \S 2 are expected to hold exactly, as long as the assumed statistical 
homogeneity and stationarity are satisfied. However, the artificial numerical diffusion of the density 
field, which is unavoidable in simulations but absent in real flows, may cause departures from the 
exact results. As the continuity equation is evolved on a discrete grid, strictly speaking, the computed 
density field is not the exact solution. For example, the length scale of density structures is 
limited by the size of the computational cell, and intense structures such as shocks would 
appear more diffuse than in real flows. This effect of discretization is responsible for the departure 
of the simulation results from our exact relations, Eqs.\ (\ref{zero-div}) and (\ref{zero-adv}), 
which may be used as a tool to evaluate the accuracy of the simulations.

The numerical diffusion in a specific simulation also depends on the adopted solver 
and the details of its implementation, such as the regularization methods used to stabilize the shocks. 
To examine the effects of numerical diffusion, we adopt a generic form, $\nabla \cdot (\kappa (\rho) \nabla \rho)$, 
where, for simplicity, the diffusivity $\kappa(\rho)$ is assumed to depend only on the density. 
This assumed form of the diffusion gives a contribution of $\kappa (\nabla s)^2 + \nabla \cdot (\kappa \nabla s)$
to Eq.\ (\ref{s-eq}), which leads to,
\begin{multline}
\frac {\partial f (s; t)} {\partial t}  =  
\frac {\partial}  {\partial s} \Big( \langle \bs{u} \cdot \nabla s |s \rangle f -  \kappa (s)\langle  (\nabla s)^2 |s \rangle f \Big)+ \\
\hspace {-5cm}\frac {\partial^2}  {\partial s^2}  \Big(\langle \bs{u} \cdot \nabla s |s  \rangle f - \kappa (s)\langle  (\nabla s)^2 |s  \rangle f \Big), 
\label{adv-diff}
\end{multline}  
which is again a Fokker-Planck equation. We will refer to 
$\kappa (s)\langle  (\nabla s)^2 |s  \rangle$ as the conditional mean dissipation of 
$s$. 
The drift and diffusion coefficients indicate a competition between the 
conditional mean advection and the numerical diffusion, which tend to broaden and narrow 
the PDF, respectively.  At steady state, the two terms must cancel out, i.e., 
$\langle \bs{u} \cdot \nabla s |s \rangle  =  \kappa(s) \langle (\nabla s)^2 |s  \rangle$, suggesting that the numerical diffusion of 
density tends to make the conditional mean advection positive rather than 0.
The relation provides an estimate of the effective numerical diffusivity as a function of $s$,
\begin{equation}
\kappa(s) = \langle \bs{u} \cdot \nabla s |s \rangle  / \langle (\nabla s)^2 |s  \rangle.
\label{kappaeq}
\end{equation}
The effect of numerical diffusion is expected to decrease with increasing resolution, 
so our exact results, Eqs.\ (\ref{zero-div}) and (\ref{zero-adv}), should be better 
satisfied at higher resolution.

\subsection{Comparison with simulation data}

We simulated an isothermal, supersonic turbulent flow with rms Mach number $\sim7.5$, using the recently 
developed code \emph{Dispatch} \citep{Nordlund+18}. We solved the 3D hydrodynamic equations without 
explicit viscosity in a periodic simulation box of unit size, using the HLLC (Harten-Lax-van Leer-Contact) approximate 
Riemann solver \citep{Toro+94}. The flow was driven with a solenoidal random force in Fourier space at wavenumbers $1<k/2\pi<2$. 
The simulations lasted 15 dynamical times, and we used the last 30 snapshots covering 10 dynamical times for  
statistical analysis. In order to examine the dependence on numerical resolution, we carried out simulations at four resolutions, from $128^3$ to $1024^3$. 
Note that our theoretical results are general, applicable to all compressible flows, and the specific choice of code and simulation here is 
intended only to show how to use our exact results to test the accuracy of simulations and to illustrate the method to estimate numerical diffusivity. 

\begin{figure}[t]
\includegraphics[width=0.9\columnwidth]{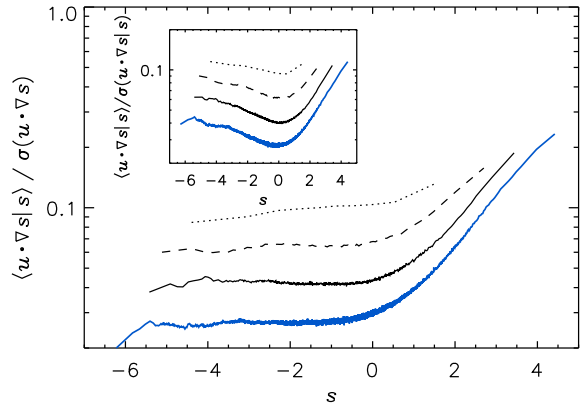}
\caption[]{Conditional mean advection, $\langle  \bs{u} \cdot  \nabla s |s \rangle$, normalized to 
the rms, $\langle  ( \bs{u} \cdot  \nabla s)^2 \rangle^{1/2}$, of the advection term. The inset shows the same quantity normalized to 
the conditional rms $\langle  ( \bs{u} \cdot  \nabla s)^2 |s \rangle^{1/2}$ in each $s$ bin. Dotted, dashed, solid  and blue solid lines show results at 
$128^3$, $256^3$, $512^3$ and $1024^3$ resolution, respectively.
}
\label{ugrads}
\end{figure}

To compute the conditional statistics, we divide the $s$ space into bins of different widths, 
such that the sample size of each bin is constant, $\simeq$ 6.3 million. We first analyze the conditional 
mean advection at steady state. Fig.\ \ref{ugrads} plots  $\langle {\bs u} \cdot  \nabla s |s\rangle$ at different resolutions. 
The main panel normalizes it to the overall rms advection in the entire flow. 
Our theory predicts that the conditional mean advection vanishes exactly.  
However, the existence of numerical diffusion causes $\langle {\bs u} \cdot  \nabla s |s\rangle$ to be 
positive at all $s$, because it is in balance with the positive-definite conditional mean dissipation $\kappa(s)\langle (\nabla s)^2|s \rangle$ at steady state 
(\S 3.1). Therefore, the departure of $\langle {\bs u} \cdot  \nabla s |s\rangle$ from zero reflects the 
amplitude of numerical diffusion of density. Fig.\ \ref{ugrads} shows that $\langle {\bs u} \cdot  \nabla s |s\rangle$ is small and 
almost constant at small $s$, and then quickly rises at $s\gsim 1$, indicating an increase of the 
numerical dissipation with $s$.

The inset of Fig.\ \ref{ugrads} shows that the ratio of the conditional mean 
advection to the conditional rms, $\langle  (\bs{u}\cdot  \nabla s)^2 |s \rangle^{1/2}$, in each $s$ bin. 
The ratio reflects how close to zero the conditional mean advection is. The ratio is much smaller than 1;  
it 
is only $ \simeq 0.12$ at the lowest resolution,  and decreases steadily with increasing resolution, to about $0.03$ for 
small $s$ at $1024^3$. This continuous decrease with increasing resolution, without any sign of convergence, is consistent with our 
theory that predicts $\langle  {\bs u} \cdot \nabla s |s\rangle=0$ in the absence of numerical diffusion, which can be 
achieved only toward infinite resolution.

\begin{figure}[t]
\includegraphics[width=0.9\columnwidth]{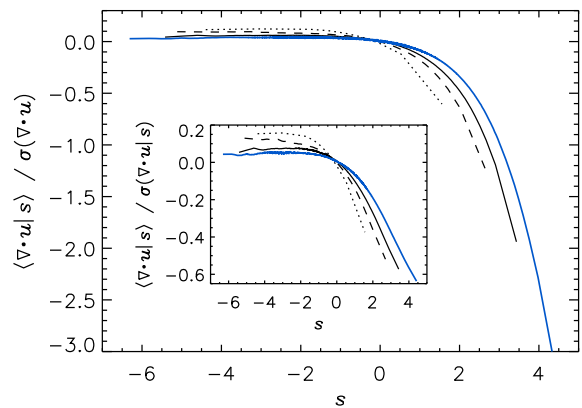}
\caption[]{Conditional mean divergence, $\langle  \nabla \cdot \bs{u}  |s \rangle$, normalized to the overall rms flow divergence, $\langle  (\nabla \cdot \bs{u})^2 \rangle^{1/2}$
(main panel), and to the conditional rms $\langle  (\nabla \cdot \bs{u})^2 |s \rangle^{1/2}$ (inset) in each $s$ bin. }
\label{div}
\end{figure}

In Figure \ref{div},  we plot the conditional mean divergence $\langle \nabla \cdot  \bs{u} |s \rangle$ measured from the simulation data. 
When normalized to the overall rms divergence, $\langle (\nabla \cdot  \bs{u})^2\rangle^{1/2}$ (main panel), 
the conditional mean divergence is close to zero and almost constant at small $s$. It then starts decreasing at $s \simeq -1$, 
and finally becomes negative and deviates significantly from zero at the largest values of $s$.  
The conditional mean divergence was also predicted to be zero, and like the case of $\langle  {\bs u} \cdot \nabla s |s\rangle$, its 
departure from zero at large $s$ also corresponds to the effect of numerical diffusion (see below). 
Intuitively, strong shocks in a simulation may result in structures that are initially unresolved, 
with lower densities than expected, and thus their negative divergence are artificially assigned to relatively 
lower densities. This contributes to the negative mean divergence, $\langle \nabla \cdot  \bs{u} |s \rangle$, at large $s$.

The inset of Fig.\ \ref{div} normalizes $\langle \nabla \cdot  \bs{u} |s \rangle$ to the conditional rms, $\langle  (\nabla \cdot \bs{u})^2 |s \rangle^{1/2}$, of the divergence.
This normalization is a better indicator of how well the negative and positive parts of the divergence PDF, which is generally 
very broad, cancel out at each given density. Except at the largest values of $s$, where the conditional mean to rms ratio reaches 
$\simeq -0.6$, $\langle \nabla \cdot  \bs{u} |s \rangle$ is significantly smaller than $\langle  (\nabla \cdot \bs{u} )^2 |s \rangle^{1/2}$, 
especially at high resolutions. 
For both normalizations, the conditional mean divergence gets closer to zero with increasing 
resolution, again with no sign of convergence at $1024^3$. 
It is thus likely that $\langle \nabla \cdot  \bs{u} |s \rangle$ continues to approach zero as the resolution increases further, 
consistent with our prediction that, in the absence of numerical diffusion, the divergence is perfectly balanced at each density level.

\begin{figure}[t]
\includegraphics[width=0.9\columnwidth]{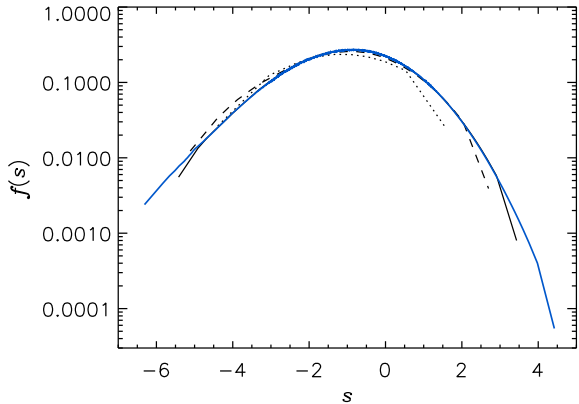}
\caption[]{Probability distribution of $s$ using the same $s$ bins as for the analysis of the conditional means.}
\label{figpdf}
\end{figure}

To understand the significant departure of $\langle \nabla \cdot  \bs{u} |s \rangle$ from 0 at large $s$, we 
make use of Eq.\ (\ref{div-adv-relation}), which provides a relation between $\langle \nabla \cdot  \bs{u} |s \rangle$ and $\langle {\bs u} \cdot  \nabla s |s\rangle$. 
We rewrite the relation as $\langle \nabla \cdot  \bs{u} |s \rangle = \partial_s \langle {\bs u} \cdot  \nabla s |s\rangle + \langle {\bs u} \cdot  \nabla s |s\rangle \partial_s [\ln f(s)]$.  
The second term, $\langle {\bs u} \cdot  \nabla s |s\rangle \partial_s \ln f(s)$, is dominant at large $s$ because 
the right tail of the density PDF, $f(s)$, decreases very fast  (see Fig.\ \ref{figpdf}). Therefore, 
the decrease of $\langle \nabla \cdot  \bs{u} |s \rangle$ below zero at large $s$ corresponds to the decrease of 
$f(s)$ and the rise of $\langle {\bs u} \cdot  \nabla s |s\rangle$ toward large densities. 
In particular, the fast decrease of $f(s)$ at large $s$ explains why the departure 
of $\langle \nabla \cdot  \bs{u} |s \rangle$ from zero is more significant than that of $\langle {\bs u} \cdot  \nabla s |s\rangle$. 
Since the increase of $\langle {\bs u} \cdot  \nabla s |s\rangle$ at 
large $s$ is caused by numerical diffusion, the significant departure of $\langle \nabla \cdot  \bs{u} |s \rangle$ from zero at large $s$ also reflects the effect 
of numerical diffusion. At small $s (\lsim -2)$, both $\langle \nabla \cdot  \bs{u} |s \rangle$ and $\langle {\bs u} \cdot  \nabla s |s\rangle$ 
appear to be roughly constant (Figs.\  \ref{ugrads} and \ref{div}). For 
constant $\langle \nabla \cdot  \bs{u} |s \rangle$ and $\langle {\bs u} \cdot  \nabla s |s\rangle$, Eq.\ (\ref{div-adv-relation}) implies $f(s)$ 
is exponential, which is approximately consistent with the left PDF tail shown in Fig.\ \ref{figpdf}. However, the approximately 
exponential left tail is not expected in general, as its shape may depend on various factors, such as the flow Mach number, 
the driving pattern \citep[]{Federrath+10}, and possibly the numerical code.

We have found that, consistent with our theory, $\langle \nabla \cdot  \bs{u} |s \rangle$ and $\langle  \bs{u} \cdot  \nabla s |s \rangle$ 
are close to 0 at small $s$, and their departure from 0, occurring primarily at large $s$, reflects the artifacts of numerical diffusion 
and steadily decreases with increasing resolution. 

\begin{figure}[t]
\includegraphics[width=0.9\columnwidth]{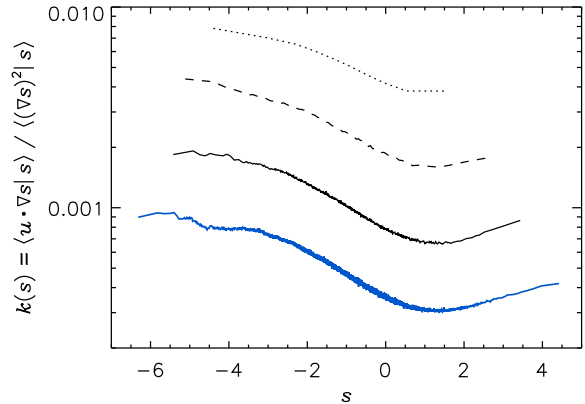}
\caption[]{Measured numerical diffusivity, $\kappa(s)$, as a function of logarithmic density at different numerical resolutions.
}
\label{kappa}
\end{figure}

Our analytical work provides a way to estimate the effective numerical diffusivity, $\kappa(s)$ (Eq.\ \ref{kappaeq}). 
As expected, Fig.\ \ref{kappa} shows the measured $\kappa(s)$ decreases with increasing resolution. 
For each factor of 2 increase in resolution, $\kappa(s)$ decreases 
by a factor of $\simeq 2$ for all $s$, and the shape of $\kappa(s)$ as a function 
of $s$ appears to be independent of the resolution. 
The invariance is likely a result of the fact that Riemann solvers resolve shocks with a 
fixed number of cells, corresponding to a diffusivity that scales linearly with the cell size.  
Finite differences solvers with diffusivities proportional to the cell size are 
expected to show similar scaling behavior.
If so, $\kappa(s)$ may be viewed as an intrinsic feature that characterizes
each code. The invariance of the function form of $\kappa(s)$ with resolution in our simulation 
may  be partly responsible for the convergence of the overall shape of the density PDF. 
As seen in Fig.\ (\ref{figpdf}), the shape of $f(s)$ is also largely invariant with numerical resolution. 
The effect of increasing resolution is mainly to extend the PDF to a wider $s$ range. 

At all resolutions, $\kappa(s)$ decreases by a factor of $\simeq 3$ as $s$ increases to 1, 
and then slightly rises as $s$ increases further. The decrease of  $\kappa(s)$ with 
increasing $s$ does not imply the numerical dissipation, $\kappa (s) \langle (\nabla s)^2|s\rangle$, is weaker at 
larger $s$. 
In fact, the larger departure from zero of  $\langle \nabla \cdot  \bs{u} |s \rangle$ and $\langle  \bs{u} \cdot  \nabla s |s \rangle$
at large $s$ is due to the increase of the numerical dissipation of $s$ toward large densities. 
Although $\kappa(s)$ at large $s$ is already 2-3 times smaller than at small $s$, it is still 
not sufficient to keep the numerical dissipation of $s$ in dense regions 
at a satisfactory level.  Adopting adaptive-mesh-refinement methods could help further 
reduce the numerical diffusion at large $s$. 

A fundamental question concerning density fluctuations in supersonic turbulence is how well the 
density PDF of a simulation represents the PDF, $f_{\rm real}(s)$, of a real flow. 
The convergence of the PDF with resolution does not necessarily 
guarantee the PDF is accurate. Numerical diffusion is unavoidable in simulations, 
and its dependence on $s$ may leave an artificial imprint on the density PDF. It is unknown 
what function form of $\kappa(s)$ would give density statistics closest to $f_{\rm real}(s)$. 
Furthermore, if the function form of $\kappa(s) $ with $s$ is invariant with resolution, as 
in our case, increasing resolution may not bring the shape of the PDF closer to 
$f_{\rm real}$, as it may only extend the PDF to a wider density range. 
According to our preliminary results (Pan  et al. 2018, in preparation), numerical diffusion in simulations 
may cause a significant underestimate of the high-density tail of the density PDF.  A semi-analytical approach 
developed in our new work that removes the direct effect of the artificial diffusion of the density field predicts 
a power-law tail, while numerical simulations always yield a nearly lognormal tail.

Based on our exact results, we propose to use the departure from zero of the conditional means, 
$\langle \nabla \cdot  \bs{u} |s \rangle$ and $\langle  \bs{u} \cdot  \nabla s |s \rangle$, at all values of $s$,
as an objective criterion to evaluate the ability of numerical codes to reproduce the correct density PDF. 
This criterion will be adopted in the context of a future systematic study of the shape of the density 
PDF in turbulent flows simulated with different codes.

\section {Conclusions} 
We have used both analytical and numerical approaches to investigate density 
fluctuations in compressible turbulence. 
Kinetic equations for the density PDF were derived in two forms, a first-order partial differential equation (Eq.\ \ref{pdf-div}) 
and a Fokker-Planck equation (Eq.\ \ref{pdf-adv}) with coefficients given by the conditional mean 
divergence, $\langle \nabla \cdot  \bs{u}  |s \rangle$, and advection, $\langle  \bs{u}  \cdot \nabla s  |s \rangle$, 
respectively. With the assumption of statistical homogeneity only, two exact results were predicted, 
$\langle \nabla \cdot  \bs{u} |s \rangle = 0$ and $\langle  \bs{u}  \cdot \nabla s  |s \rangle =0$, 
corresponding to the detailed balance of probability fluxes at steady state. 
In simulations, the departure of the conditional mean divergence and advection from 0 corresponds to the artifacts of the numerical diffusion 
of the density field. Our exact results provide an objective measure for the accuracy of the density PDF from numerical simulations, 
suggesting that the codes yielding smaller departure of $\langle \nabla \cdot  \bs{u} |s \rangle$ and $\langle  \bs{u}  \cdot \nabla s  |s \rangle$ from 0 are 
to be considered more accurate. A general method is also developed 
to measure the numerical diffusivity, $\kappa(s)$, as a function of $s$, which may be used to characterize 
each numerical code. 
A systematic study of the effects of numerical diffusion on the PDF shape using different 
codes is being planned and will be reported in future work. 

\acknowledgements
LP acknowledges support from the Youth Program of the Thousand Talents Plan in China.
PP acknowledges support  by the Spanish MINECO under project AYA2017-88754-P (AEI/FEDER,UE). 
The work of {\AA}N was supported by grant 1323-00199B from the Danish Council for Independent 
Research (DFF). The Centre for Star and Planet Formation is funded by the Danish National Research Foundation (DNRF97).
Storage and computing resources at the University of Copenhagen
HPC centre, funded in part by Villum Fonden (VKR023406), were used to carry out the 
simulations.

\end{document}